\documentclass[fleqn,usenatbib]{mnras}

\usepackage{newtxtext,newtxmath}

\usepackage[T1]{fontenc}

\DeclareRobustCommand{\VAN}[3]{#2}
\let\VANthebibliography\thebibliography
\def\thebibliography{\DeclareRobustCommand{\VAN}[3]{##3}\VANthebibliography}

\usepackage{graphicx}	
\usepackage{amsmath}	

\title[Winds in Accretion Disc Coronae]{A different view of wind in X-ray binaries: The Accretion Disc Corona source 2S 0921-630}

\author[R. Tomaru et al.]{
Ryota Tomaru,$^{1}$\thanks{E-mail: ryota.tomaru@durham.ac.uk}
Chris Done,$^{1,2}$
Hirokazu Odaka$^{3,2}$
Atsushi Tanimoto$^{4}$
\\
$^{1}$Centre for Extragalactic Astronomy, Department of Physics, Durham University, South Road, Durham DH1 3LE, UK \\
$^{2}$Kavli Institute for the Physics and Mathematics of the Universe (IPMU), The University of Tokyo, 5-1-5 Kashiwa,  Chiba 277–8583, Japan\\
$^{3}$ Department of Earth and Space Science, Graduate School of Science, Osaka University, 1-1 Machikaneyama, Toyonaka, Osaka 560-0043, Japan\\
$^{4}$ Kagoshima University, Graduate School of Science and Engineering, Kagoshima 890-0065, Japan
}

\date{Accepted XXX. Received YYY; in original form ZZZ}

\pubyear{2023}

\begin{document}
\label{firstpage}
\pagerange{\pageref{firstpage}--\pageref{lastpage}}
\maketitle

\begin{abstract}

Accretion disc coronae (ADC) sources are very high inclination neutron star or black hole binaries, where the outer accretion flow blocks a direct view of the central source.
The weak observed X-ray emission is instead produced mainly by scattering of the intrinsic radiation from highly ionised gas surrounding the source, the ADC.
However, the origin of this scattering material is still under debate.
We use the ADC source 2S 0921-630 (V395 Car) to test whether it is consistent with a thermal-radiative wind produced by the central X-ray source illuminating and puffing up the outer disc.
This wind is clearly visible in blueshifted absorption lines in less highly inclined systems, where the source is seen directly through this material.
Using the phenomenological photoionised plasma model, we first characterise the parameter that drives emission lines observed in 2S0921 in {\it  XMM-Newton} and {\it Chandra} data.
Following this, we run the Monte Carlo radiation transfer simulation to get scattered/reprocessed emissions in the wind, with the density and velocity structure obtained from the previous work.
Our model agrees with all the wind emission lines in the {\it Chandra} high and medium energy grating spectra for an intrinsic source luminosity of $L > 0.2~L_{\rm Edd}$. 
This result strongly favours thermal-radiative winds as the origin of the ADC.
We also show how high-resolution spectra via microcalorimeters can provide a definitive test by detecting blueshifted absorption lines.
\end{abstract}




\section{Introduction}

The nature and geometry of the accretion flow in a strong gravity field are still not well known.
This is the most obvious case for the central regions, where the X-ray flux is emitted, 
but even the outer accretion disc structure is uncertain and affected by the central regions. 
X-ray irradiation from the strong central source
results in an X-ray heated and ionised photosphere \citep{Cunningham1976, Vrtilek1996}, so the outer disc subtends a larger solid angle than expected from an unirradiated disc model \citep{Jimenez-Garate2002}. Further vertical extent arises if there is a wind from the disc.
Physically, winds are expected when the Compton temperature of the X-ray heated material, $T_{\rm IC}$,
gives typical particle velocities which can overcome the local gravity \citep{Begelman1983a}. The density and velocity structure of these winds can be determined uniquely from 
the intrinsic spectrum and luminosity of the central source, together with a prescription for the disc geometry.
Such thermal winds (helped by radiation pressure when the source luminosity $\ge 0.1L_{\rm Edd}$) from the outer disc well describe the mildly blue-shifted (velocity of $100-1000~ \rm{km s^{-1}}$), highly ionised absorption features seen from high inclination binary systems with large discs (see, e.g. \citealt{Tomaru2019, Tomaru2020, Tomaru2020b}).
However, absorption lines  only give information on the line of sight density/velocity structure, so they cannot show the full 2-dimensional extent of the wind.

Emission lines have contributions from the full extent of the wind flow.
The amount of emission compared to absorption (P Cygni profile) traces the solid angle subtended by the wind, while the 
intrinsic width of the line is determined by the integrated velocity structure across all azimuths of the wind. 
Thus the emission lines give additional information about the wind structure, geometry and energetics, complementary to the absorption lines. 
However, the emission lines are very difficult to see in current data, as they have fairly small equivalent widths against the continuum, and fairly small intrinsic widths, so they cannot be resolved even with High-Energy Transmission Gratings Spectrometer (HETGS) onboard {\it Chandra}.

Instead, in the Accretion Disc Corona (ADC) sources, the direct source emission is blocked by the vertically extended disc and its atmosphere 
due to the extremely high inclination angle \citep{White1982b, Hellier1989, Kubota2018}.
The observed X-ray continuum comes from the small fraction of the central source scattered in material above the disc (the ADC),
so the emission lines from this same material become prominent compared with the continuum.

Of all the known ADC sources, 2S 0921-630 (V395-Car) has the longest orbital period of 9 days with a high inclination $i\sim 82^\circ$\citep{Ashcraft2012}.
This means that it has the largest radially extended disc and also that the companion star must be significantly evolved to fill its Roche lobe at this binary separation. A high mass transfer rate and hence high luminosity is thus expected.
This combination of high luminosity and extended disc means it is the one most likely to show a strong thermal-radiative wind, and indeed {\it Chandra}/HETGS data shows  emission lines from the same highly ionised ions (e.g. Fe\,{\sc xxv, xxvi}) which are seen in  absorption in less inclined ionised, as well as smaller features from other highly ionised, lower Z elements such as S, Si and O \citep{Kallman2003}. 

Here we explore how the current {\it Chandra}/HETGS data can constrain the properties of the scattering/emission region in this source and compare it to the detailed line profiles predicted from thermal-radiative wind models.
These are built using state-of-the-art radiation hydrodynamic models to derive the density and velocity structure of the thermal-radiative wind,
together with Monte-Carlo radiation transfer to determine the line profiles in the non-spherical geometry.
We show that our wind emission model describes the current data well. 
We show that higher resolution data from the microcalorimeter (Resolve) onboard the next X-ray satellite, X-Ray Imaging and Spectroscopy Mission ({\it XRISM}, \citealt{Tashiro2020})  will  allow the velocity structure to be explored in both emission and absorption. 
Direct detection of a blue-shifted absorption line will unambiguously show that this scattering material is a wind rather than an atmosphere above the disc. 

\section{System parameters of 2S 0921-630}

The orbital period is 9 days and the masses of the binary components are $M_{\rm NS}= 1.4 M_\odot$, $M_{s} = 0.4 M_\odot$ \citep{Steeghs2007}.
The outer disc radius can be determined by assuming that the 
disc extends to the tidal radius which is (approximately) 80 percent of the Roche-lobe radius.
This gives $R_{\rm disc}=1.4\times 10^{12} {\rm cm}$.

\if
$ = 10 R_{\rm IC,7} $,
where the $R_{\rm IC ,7}= 0.61 m_p G M_{\rm NS}/(kT_{\rm IC, 7})$ and $T_{\rm IC, 7} = T_{\rm IC}/(10^{7}~{\rm K})$. 
\fi
\if
We use the bright NS LMXRB system GX 13+1 as a comparison source. This is a non-ADC system, where the central X-ray source can be seen directly, with multiple strong absorption lines showing that the line of sight is at a high enough inclination to intercept the wind. GX 13+1 has a longer orbital period at 24 days, but its companion is a K giant with mass $\sim 5 M_\odot$ so the outer disc radius is comparable at $\sim 10^{12}$~cm. 
\fi

There is some discussion in the literature on whether S0921-630 really is an ADC source (see e.g. \citealt{Kallman2003}), as the distance and hence luminosity were quite uncertain.
However, there is now a parallax measurement from {\it GAIA} DR3 of $0.0607\pm	0.0231$ milliarcsec. 
Simply inverting the parallax gives a poor estimator of distance where  uncertainties are large, so we unfold against a stellar distribution prior to get the full Bayesian posterior \citep{Bailer-Jones15}.
This gives a maximum probability distance of 9.1~kpc (90\% confidence interval 6.8--13.1~kpc, Y. Zhao, private communication).
Neutron star spectra show a transition from hard to soft states at $0.01-0.1_{\rm Edd}$ or $L\sim 2-20\times 10^{36}$~ergs~s$^{-1}$ (e.g. \citealt{Gladstone2007}).
The broadband (0.5-25~keV) spectrum of 2S0921 as seen in {\it Suzaku} clearly shows the rollover shape expected from a soft state but at an observed flux of $\sim 10^{-10}$~ergs~cm$^{-2}$~s$^{-1}$ (\citealt{Sharma2022, Yoneyama2022}).
This corresponds to $L\sim 10^{36}$~ergs~s$^{-1}$, below the threshold for the soft state if seen directly. 
Additionally, the observed X-ray spectrum contains multiple high ionisation emission lines, which is not usual for low-mass X-ray binaries seen directly but is expected from a source spectrum dominated by scattering in diffuse gas above the disc.
Hence it seems most likely that this source is an ADC, and we assume this in the remainder of the paper.

\section{Data analysis}

We analyze the broadband spectrum of the {\it XMM-Newton}/EPIC-PN (obsID 0051590101, 20 Dec 2000) together with the higher resolution grating data from {\it Chandra}/HETGS (obsID 1909, 2 Aug 2001). The same data sets are used by \cite{Kallman2003}. 
We first explore these data by characterising the broadband continuum and then use phenomenological photoionised plasma models to characterise the observed emission lines.

We use the {\it XMM-Newton}/EPIC-PN pipeline products (data, background and ancilary response file) provided in the archive without further processing. 
We obtain the {\it Chandra}/HETGS data via Chandra Interactive Analysis of Observations ({\sc ciao}) v4.9.
We combine positive and negative 1st-order spectra of HEG and MEG using 
{\tt combine\_grating\_spectra}, and rebin
both spectra using GRPPHA in FTOOLS to have at 20 counts ${\rm bin^{-1}}$.
The {\it XMM-Newton} and {\it Chandra} data are not simultaneous but show similar spectral shapes and are within a factor of 2 of the same flux. 


\begin{figure}
    \centering
    \includegraphics[width=\hsize]{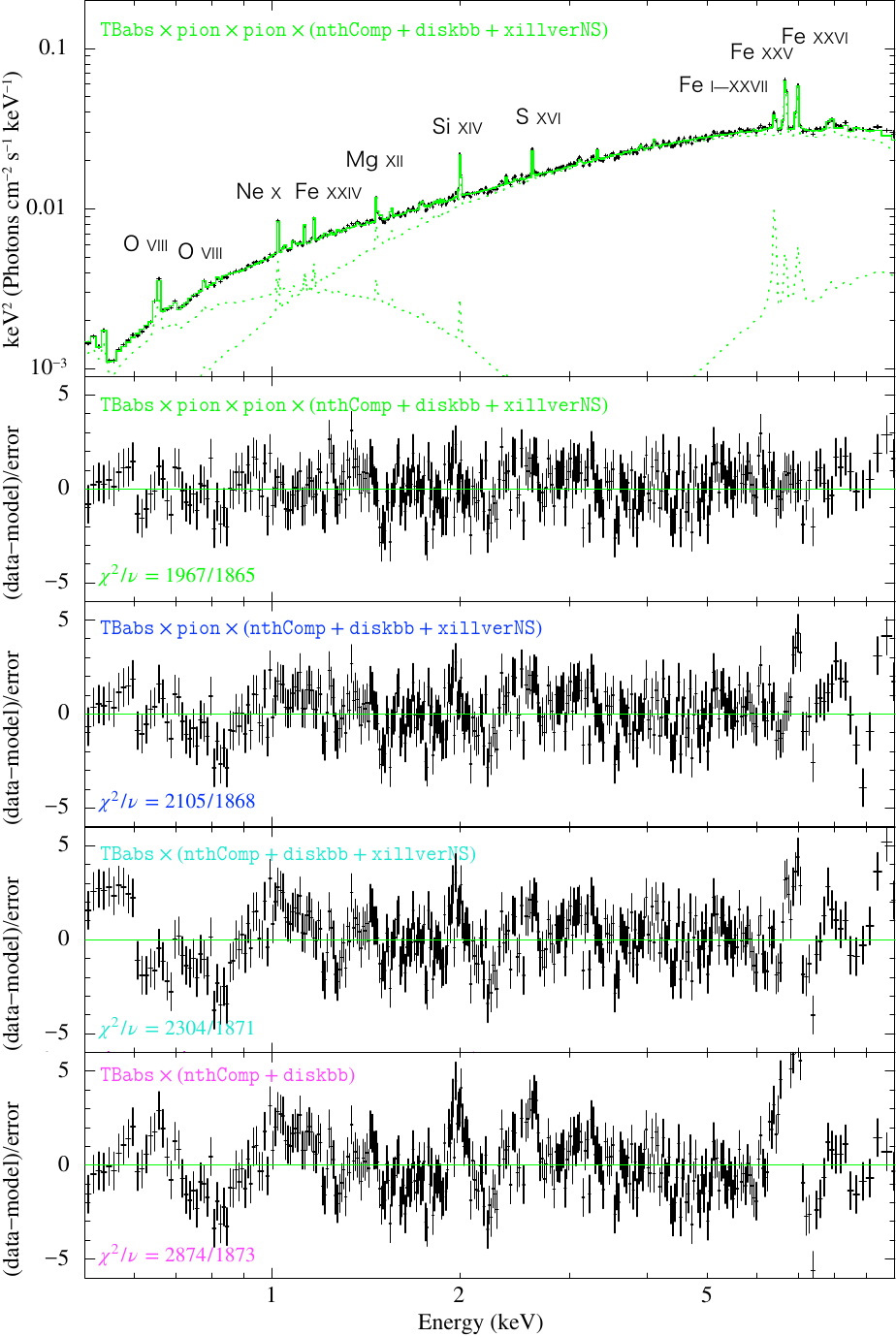}
    \caption{The best fit broad band spectrum of {\it XMM}/EPIC-PN with two {\sc pion} photoionised emission components, together with disc reflection {\sc xillverNS}. Residuals are shown in the lower panels, first to a pure continuum model, then including the reflection component, then including a single {\tt Pion}, and then the full model. Parameters for all the fits are in Table~\ref{tab:fit_xmm}. 
    Note that these figures are rebinned for plotting purposes only. 
    }
    \label{fig:xmm}
\end{figure}

\subsection{Characterising the spectrum from XMM-Newton broadband data}

The standard soft state continuum in neutron star binaries can be fitted by a disc and higher temperature boundary layer.
The boundary layer in bright neutron shows 
a fairly constant spectral shape, 
similar to a blackbody with temperature of  $\sim 2.4$~keV but somewhat broader, showing that it is Comptonised \citep{Revnivtsev2006}.
Thus we first assume that the continuum spectrum is the sum of disc blackbody ({\tt diskbb}, \citealt{Mitsuda1984, Makishima1986}) and thermal Comptonisation ( {\tt nthcomp}, \citealt{Zdziarski1996, Zycki1999}) in {\sc xspec}.
We allow the seed photon temperature in the Comptonisation component to be different to that of the disc as it is more likely that these are dominated by the neutron star surface (see also \citealt{Gierlinski2002}).

\begin{figure*}
    \centering
    \includegraphics[width=\hsize]{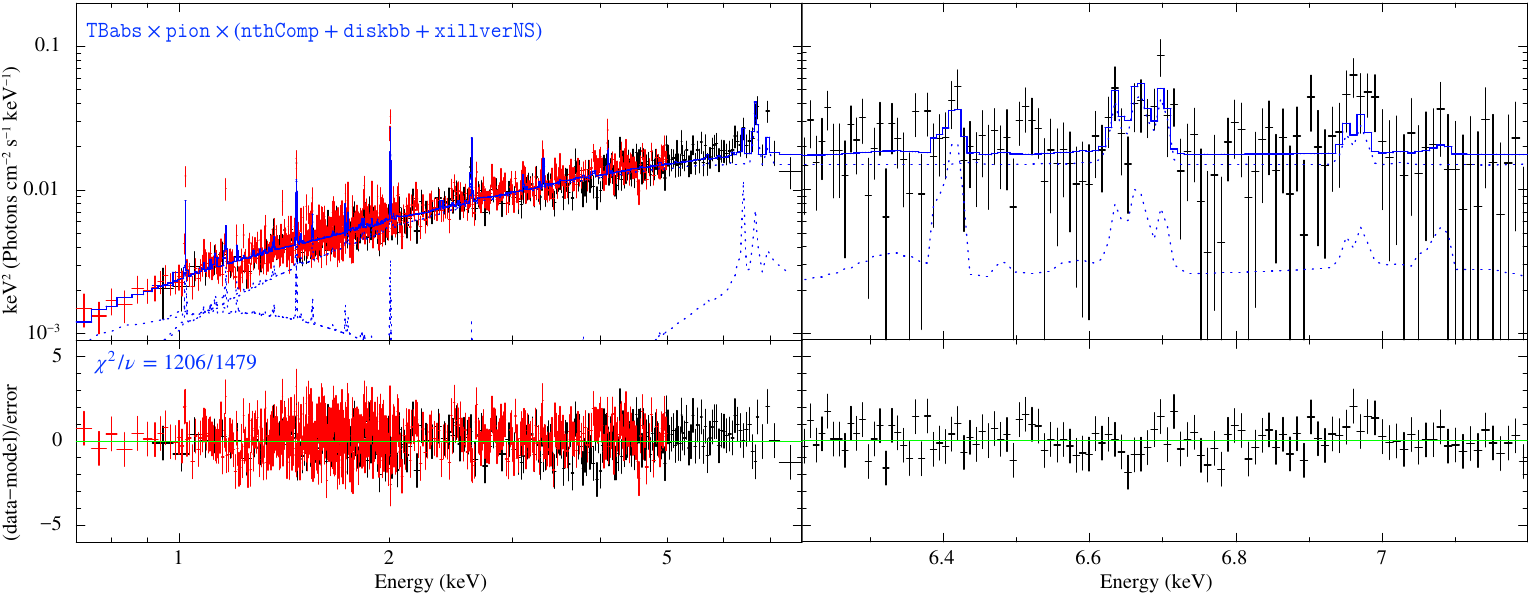}
    \caption{The left panel shows the broad-band spectrum of {\it Chandra}/HETGS fit with the continuum model plus  single {\tt Pion} emission component and {\sc xillverNS} reflection. 
    The colours show the spectrum of HEG (black), MEG(red), and the model (blue).
    The right panel shows the  unbinned narrow-band spectrum around Fe (6.2-7.2 keV), showing that the Fe {\sc xxvi} line is underpredicted. 
     Note that the broadband spectrum (left) has been grouped into 20 count bins $^{-1}$ and then further rebinned for the plotting purpose.
    }
    \label{fig:heg_meg_1zone}
\end{figure*}

\begin{figure*}
    \centering
    \includegraphics[width=\hsize]{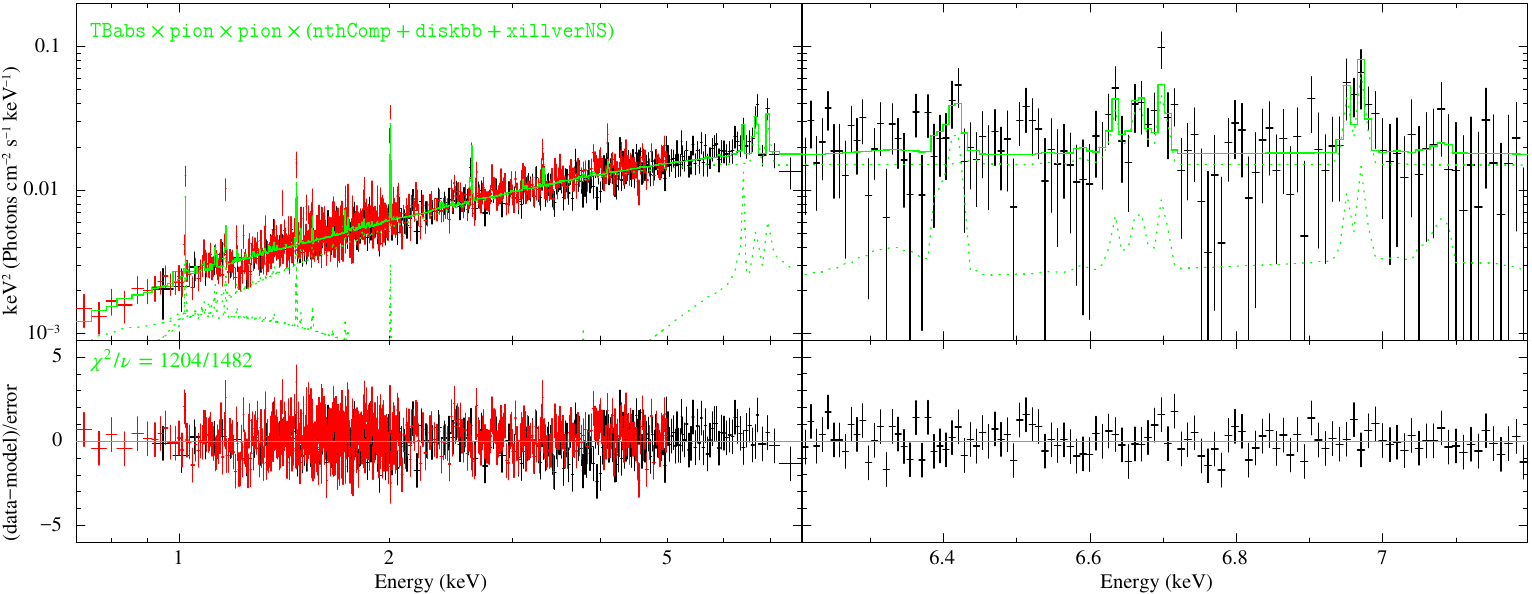}
    \caption{As in Fig.~\ref{fig:heg_meg_1zone} but with a second {\tt Pion} photoionised emission component. 
    The right panel clearly shows that this improves the fits around the Fe {\sc xxvi} line.   
    }
    \label{fig:heg_meg_2zone}
\end{figure*}

This continuum model with Galactic absorption, {\tt TBabs(diskbb+nthcomp)}, describes the shape of the broad-band continuum. The derived parameters (see Tab.\ref{tab:fit_xmm}) are fairly typical for the soft state, though the normalisation of the disc component is relatively low compared to the boundary layer emission. The 
bottom panel in Fig. \ref{fig:xmm} shows that there are strong residuals from this continuum-only model at emission line energies, as expected. The strongest lines are at 6.4-7~keV, corresponding to iron K shell transitions. 
 
Part of this iron emission can be produced by continuum reflection from the outer disc, so we model this using {\sc xillverNS} \citep{Garcia2022}, which assumes the incident spectrum is a blackbody. 
We tie this temperature to the electron temperature of the Comptonisation component {\tt nthcomp}.
The presence of reflection improves the fit but still leaves residuals (second lowest panel residuals in Fig. \ref{fig:xmm}), especially highly ionised iron (6.7-7~keV) and silicon (1.8-2~keV).

We next include emission from a photoionised plasma.
This was also included in the study of \cite{Kallman2003}, but here we use a different state-of-the-art photoionised plasma model code {\tt Pion} \citep{Mao2017} included in the {\sc spex} X-ray analysis software (v3.6, \citealt{Kaastra1996}). 
We calculate the emission lines and make a multiplicative tabular model for fitting in {\sc xspec} \citep{Arnaud1996} using a similar spectral shape for the illuminating continuum (disk blackbody {\tt DBB} and thermal Comptonisation {\tt Comt} in {\sc spex}) derived from the {\it XMM-Newton}/EPIC-PN fits. 
We fix the density at $\log (n_p/[{\rm cm^{-3}}]) =12 $, and assume a covering factor  of the photoionised region of $\Omega/4\pi=1$. 
This tabular model then has 3 free parameters, the ionisation parameter $\log (\xi/[{\rm erg~cm~s^{-1}}])$, column density $N_H [{\rm cm^{-2}}]$, and the microscopic turbulence $v_{\rm mic} [{\rm km~s^{-1}}]$. 
We include this {\tt Pion} emission model on the best-fit continuum plus reflection model and get a much better fit
(third residual from the bottom in Fig.\ref{fig:xmm}) for a column of $N_H=6\times 10^{23}$~cm$^{-2}$ at $\log\xi=3.9$.
Full fitting parameters are given in the Appendix Table\ref{tab:fit_xmm}, but we notice that the reflected spectrum now pegs to the lowest ionisation state of $\log\xi=2$.
However, there  is still an excess at $\sim 7$~keV which indicates that the H-like Fe is underestimated with a single emission component.

We add a second emission component in order to fully describe the spectra (top panel and top residual panel in Fig. \ref{fig:xmm}). 
The fit then shifts to a slightly larger column of more highly ionised material $N_H>7\times 10^{23}$~cm$^{-2}$ at $\log\xi=4.7$ together with a much lower column of less ionised material $N_H=0.8\times 10^{23}$~cm$^{-2}$ at $\log\xi=3.3$. We note that the model for the emission lines is multiplicative, so the lines appear on all the separate spectral components in the figure (including reflection).

We also checked that this model  describes the higher resolution 0.5-2~keV RGS 
spectrum taken simultaneously with the EPIC-PN data. The detailed parameters for the joint fit are consistent with those derived from EPIC-PN fitting alone, and there are no obvious line residuals remaining in the RGS data.

In our fitting, the relativistic broadening for 6.4~keV line is not required.
\citet{Yoneyama2022} explain the line width obtained by {\it Suzaku} CCDs by 
relativistic broadening, so required that the reflected component came from the inner disc, at $\sim 10^9 {\rm cm}$, while \citet{Kallman2003} instead suggest that there are multiple semi-neutral lines from the reflecting material which blend together to produce apparent broadening.
Future high resolution observations e.g with {\it XRISM} should be able to distinguish between these two scenarios for the iron line 
(see Section 6). 


\begin{figure*}
    \centering
    \includegraphics[width=\hsize]{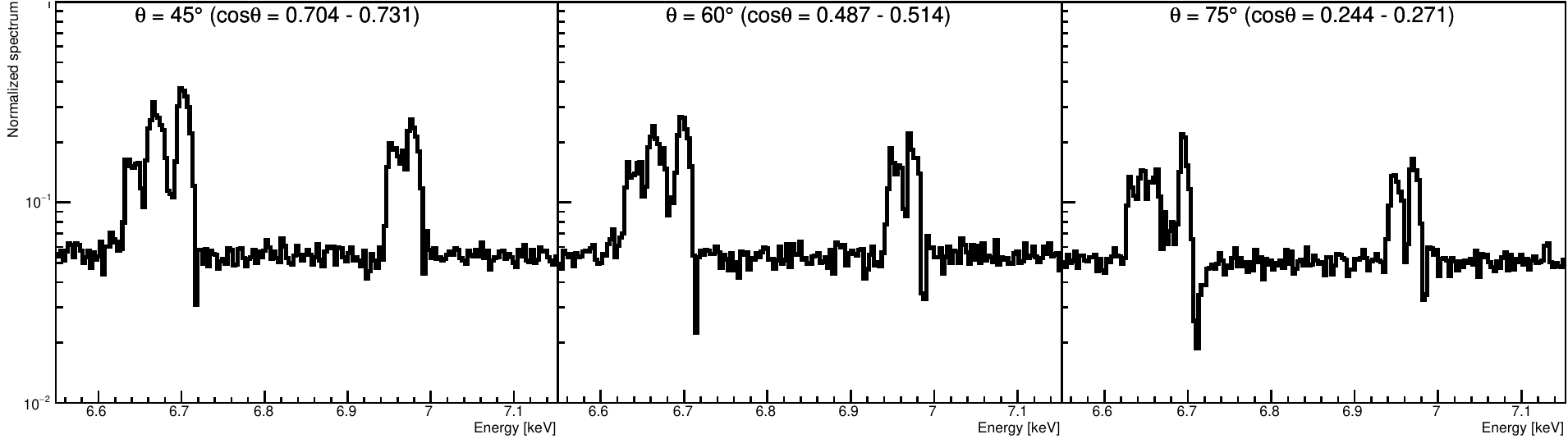}
    \caption{Results of {\sc monaco} calculation of the diffuse emission from the thermal-radiative wind seen from different inclination angles, normalised to the illuminating spectrum. The scattered continuum flux is always around 5\% of the intrinsic flux, with the emission lines superimposed. The diffuse emission from the wind is also absorbed by the wind along the line of sight at high inclination angles, as seen by the increasing strength of the absorption lines of H- and especially He-like iron  K$\alpha$ lines. 
}
    \label{fig:model}
\end{figure*}

\subsection{Chandra HETGS data}

We use the energy range of $0.9-8.0~{\rm keV}$ for the HEG 
(black in Fig.\ref{fig:heg_meg_1zone}) 
and $0.7-5.0~{\rm keV}$ for the MEG
(red in Fig.\ref{fig:heg_meg_1zone}). 

We start by using the model with one photoionised emission component, 
assuming that the continuum shape is the same as the spectrum of {\it XMM-Newton}/EPIC-PN (Fig.\ref{fig:xmm}, Tab.\ref{tab:fit_xmm} ), allowing only the normalisation to be free. 
This model fits the {\it Chandra}/HETGS data well (left in Fig. \ref{fig:heg_meg_1zone}).
The semi-neutral iron line core is now seen more clearly, and is narrow, but the {\it Chandra} data are not good enough to rule out broadening at the level suggested in the CCD data.

The parameters of the photoionised plasma model are consistent  within the error of those derived for the {\it XMM-Newton}/EPIC-PN (Tab. \ref{tab:fit_chandra}),
despite the continuum flux level being lower by factor 2.
However, again the single zone model underestimates the emission lines from Fe {\sc xxvi}, as shown by the zoom in panel (right in Fig. \ref{fig:heg_meg_1zone}).
This figure shows the unbinned 6.2--7.2 keV spectrum of HEG with the same model and parameters  as that for the broadband spectra of HEG+MEG.
Hence we also try the two-zone emission model with the same parameters as EPIC-PN except for normalization.
Although the fit statistic is almost identical, the 2-zone model clearly recovers the emission from Fe {\sc xxvi} (Fig. \ref{fig:heg_meg_2zone}). 

Thus both moderate resolution {\it XMM-Newton}/EPIC-PN and high resolution {\it Chandra}/HETGS spectra show that there is a distinct, fairly low ionisation reflector, and then additional emission lines produced from  
material with a range of higher ionisation parameters. 

\begin{figure*}
    \centering
    \includegraphics[width=\hsize]{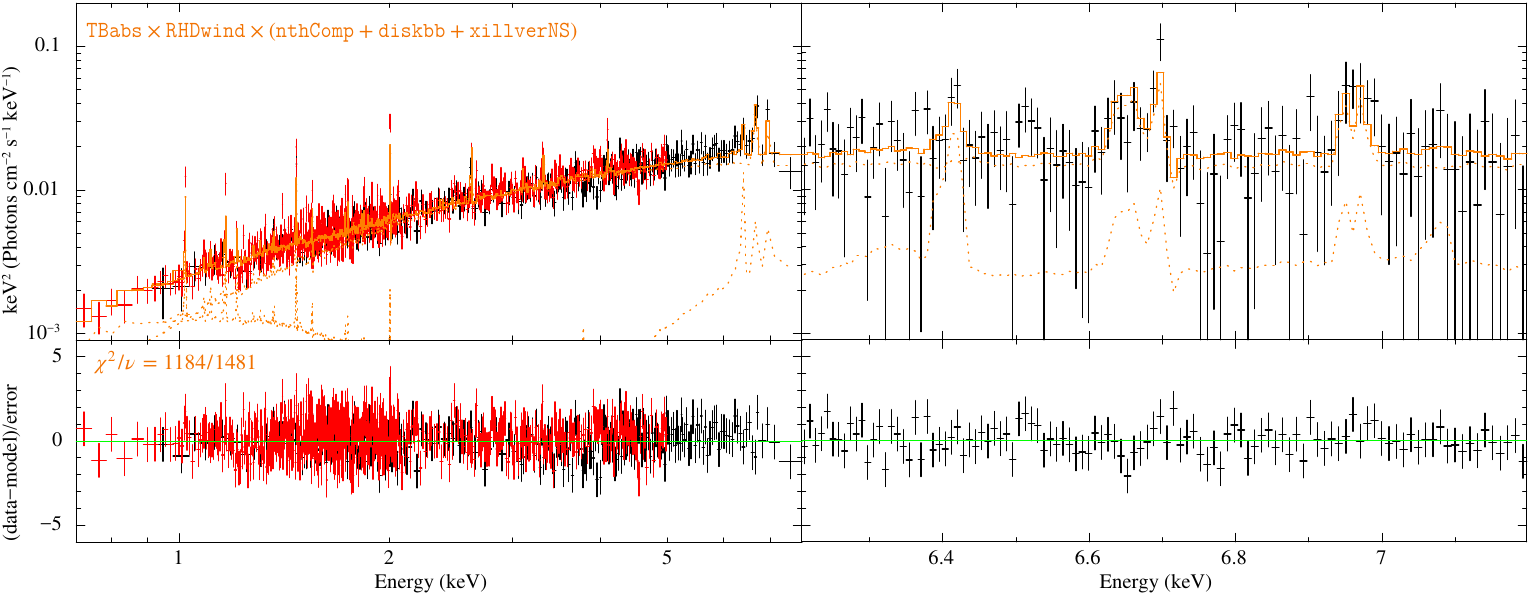}
    \caption{ {\it Chandra}/HETGS spectrum as in Fig.~\ref{fig:heg_meg_2zone} but with the two {\tt Pion} components replaced by the emission from the thermal-radiative wind simulation spectrum. The thermal-radiative wind reproduces the range of ionisation states seen in the data around He--like and H--like iron.}
    \label{fig:heg_meg_rhd}
\end{figure*}

\begin{figure*}
    \centering
    \includegraphics[width=\hsize]{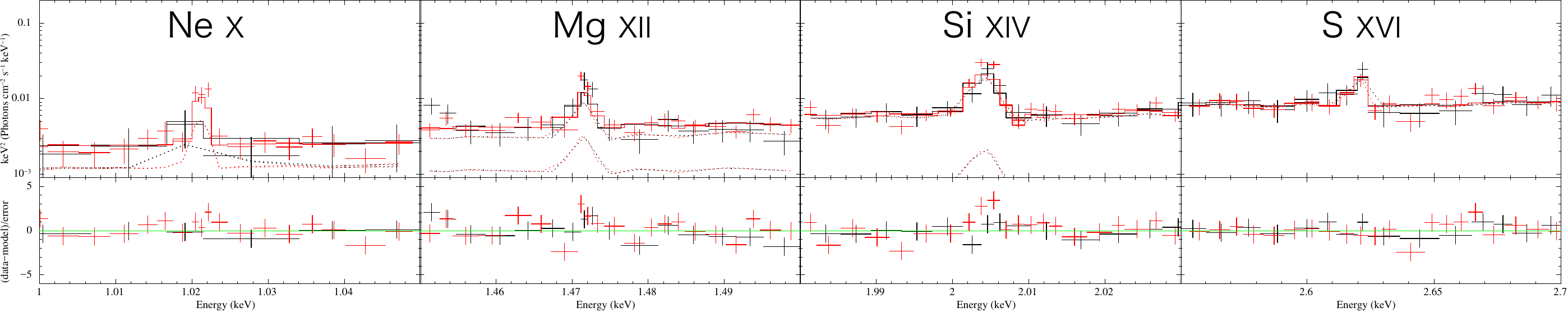}
    \caption{The fit of Fig.~\ref{fig:heg_meg_rhd} zoomed in around all the other significant emission lines. The thermal-radiative wind well reproduces the overall strength of all these features. }
    \label{fig:rhd_zoom}
\end{figure*}

\section{Predictions of a thermal-radiative wind }

We now explore whether the higher ionisation material can be produced by a thermal-radiative wind in this source. 
Such wind models can be calculated {\it ab initio} from radiation hydrodynamic codes, using the 
intrinsic spectrum and luminosity of the central source
as the input illumination of the disc whose size is derived from the binary orbit (e.g. \citealt{Woods1996,Higginbottom2014,Tomaru2019,Tomaru2020b}). However, here we do not know the intrinsic luminosity as we do not have a direct view of the central regions. 

We pick the bright LMXRB GX 13+1 as a comparable system, as it is also a neutron star with similarly large outer disc size ($R_{\rm out} = 10^{12}$~cm) so requires a similarly evolved companion star giving a similarly large intrinsic mass accretion rate.
Unlike 2S0921, in GX 13+1 we have a direct view of the central regions, so we know $L/L_{\rm Edd} =0.5$ as well as $T_{\rm IC,7} = 1.3$. 
We have already calculated the wind density and velocity structure from the radiation hydrodynamic code for GX 13+1 with these parameters, and shown detailed radiation transport using the Monte Carlo Radiation Transfer (MCRT) code ({\sc monaco} \citealt{Odaka2011}) 
through this density and velocity wind structure gives a good match to the high ionisation iron absorption lines seen from this source \citep{Tomaru2020b}.

We rerun {\sc monaco} using the same wind density, velocity and ionisation structure as in GX 13+1 but extend the energy range from 0.1-10~keV. The {\sc monaco} database includes transitions from H-, He- and Li- like
C, N, O Ne, Mg, Si, S, Ar, Ca, Mn and Cr as well as Fe and Ni, but it cannot yet handle lower ionisation species. Thus it cannot probe lines of sight
where strong low ionisation absorption should be present i.e. it cannot directly calculate the 
interface between the upper layers of the disc atmosphere/base of the wind.  
We assume here that the outer disc completely blocks our view of the central source in 2S0921 so we do not have any transmitted component through this low ionisation material in order to explore the wind emission/scattered spectrum. 
However, the reflected semi-neutral line should also be produced from this disk/wind interface region so we cannot yet calculate this self-consistently from {\sc monaco} alone. 

Fig.\ref{fig:model} shows the ratio of the
predicted wind emission/scattered flux compared to the incident photons as a function of angle in the 6.55-7.15~keV energy band. 
The scattered fraction is almost constant at $\sim 5\% $ of the intrinsic photons irrespective of inclination angle (Fig.\ref{fig:model}).
The emission line intensities are also similar for all inclination angles, but the absorption features are more significant at higher inclination angles due to the larger column density along the line of sight through the wind. 

We converted the {\sc monaco} output into an {\sc xspec} multiplicative (mtable) model to be able to compare directly to the data. Fig.~\ref{fig:heg_meg_rhd} shows this compared to the {\it Chandra}/HETGS data, with zoom-in to the iron line region (as in Sec.2.2), where the semi-neutral reflected line is modelled using {\sc xillverNS} as before.
Fig~\ref{fig:rhd_zoom} shows a zoom-in of the model fit to 
all other significant emission lines. 
The fit is qualitatively very good across the entire spectrum and across all the emission lines in the full 0.7-8~keV bandpass. 
This shows that the continuum and high ionisation lines are consistent with being produced by the wind alone, as assumed.

The {\it XMM-Newton} RGS data allow us also to explore the 
lower energy O {\sc viii} and N {\sc vii} emission lines ($\sim 0.65$ and $0.5$~keV). 
Unlike the other lines, these are overestimated by about a factor of 2 in the thermal-radiative wind model output.
This may be due to absorption from lower ionisation states in the 
region close to the disc/wind interface as the O {\sc viii} and N {\sc vii} emission lines are produced close to this transition.
This region is also the place where the disc reflection component is produced, so we cannot reliably predict the iron K$\alpha$ line from the outer disc in our current MCRT code.
We will explore how to extend the code to incorporate this complex transition region in future work.


\begin{figure*}
    \centering
    \includegraphics[width=\hsize]{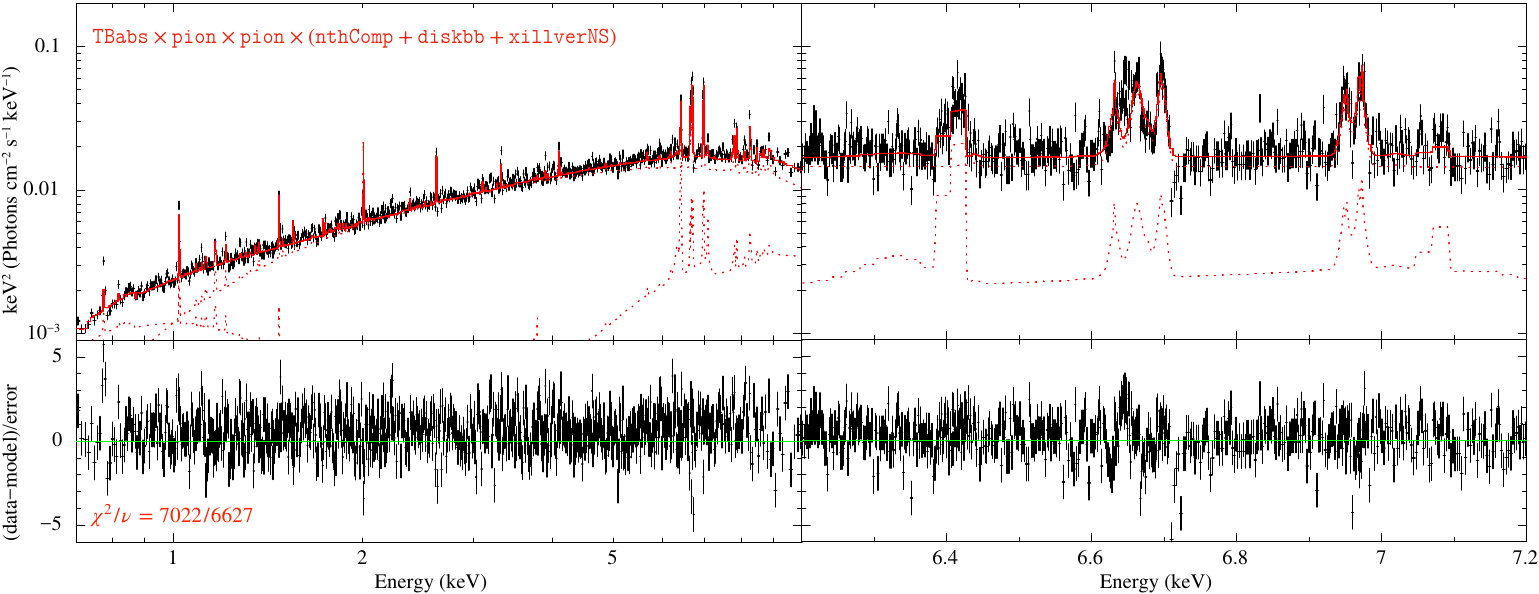}
    \caption{{\it XRISM}/Resolve calorimeter spectrum for this object simulated 
    using the thermal-radiative wind emission but fit with two {\tt Pion} photoionised emission components. There is a clear absorption residual on the blue wing of the Fe {\sc xxv} line at 6.7~keV.
    This clearly reveals the dynamical signature of the wind.
    Note that the broadband spectrum (left) is rebinned for plotting purposes only. }
    \label{fig:xrism}
\end{figure*}

\section{Intrinsic luminosity}

We can estimate the intrinsic luminosity from the scattering level around $5\%$ of the intrinsic continuum in the thermal-radiative wind. 
This fraction is the upper limit for observation  since the vertically extended disc blocks even scattered photons in reality.
This fraction is determined by the column density of the wind and its solid angle, which together are proportional to the mass loss rate.
This, in turn, is  proportional to the combination of the luminosity of the central source and the logarithmic of the disc size in units of the characteristic wind launch radius of $0.2R_{\rm IC}$ where $R_{\rm IC}=0.61m_p GM_{\rm NS}/(kT_{\rm IC,7})$ with $T_{\rm IC,7}=T_{\rm IC}/(10^7~K)$
\citep{Done2018, Hori2018, Tomaru2020b}.
Hence the scattered flux is 
$L_{\rm scatt}/L_{\rm 0} ={\rm const} \times L_{\rm 0} \log (R_{\rm disc}/(0.2 R_{\rm IC}))$

For GX 13+1 the Compton temperature of the observed spectrum is $T_{\rm IC,7}$=1.3, and $R_{\rm disc}=10^{12}$~cm so $R_{\rm disc}/R_{\rm IC}=10$. This determines the  
constant value relating the scattered fraction to the system parameters as 
$0.05 = {\rm const}\times 0.5 L_{\rm Edd} \log(50) $.
This becomes  $L_{\rm scatt}/L_{\rm Edd} = 6\times 10^{-2} \log (R_{\rm disc}/( 0.2 R_{\rm IC})) (L_{0}/L_{\rm Edd})^{2}$.

We can use this constant to scale the simulation to the parameters of 2S0921. 
The disc size is slightly larger at $R_{\rm disc}=1.4\times 10^{12}$~cm, and the inverse Compton temperature is slightly higher at $T_{\rm IC,7}$=1.7, giving $R_{\rm disc}/R_{\rm IC}=17$. 
The observed luminosity in 2S0921 is $L_{\rm scatt}/L_{\rm Edd} \sim 3\times 10^{-3} $ 
so this gives an estimate for the  intrinsic luminosity of $L_{0}/L_{\rm Edd} \sim  0.2$. This is somewhat lower than the 
intrinsic luminosity of GX 13+1, but could be an underestimate if
some of the scattered photons close to the disc are blocked by the disc atmosphere/dense wind base. 
However, the continuum spectrum of 2S0921 is somewhat harder than that of GX 13+1, which is consistent with an intrinsically less luminous source.

Unlike the analysis of \citet{Kallman2003}, we find that this intrinsic luminosity $L_{\rm X}\sim 3\times 10^{37}$~ergs~s$^{-1}$ is easily consistent with the lowest ionisation material seen in 
data. The lowest ionisation parameter in the photo-ionised emission lines from the ADC of  $\log\xi\sim 3.3$
matches to the predicted wind density of 
$\sim 10^{12}$~cm$^{-3}$ at $R>R_{\rm IC}\sim 10^{11}$~cm \citep{tomaru2023}, and the reflected iron K$\alpha$ line has $\log\xi\sim 2$ which is the expected ionisation parameter at the base of the wind launch region on the disc (see appendix of \citealt{tomaru2023}). 
However, it is not at all clear why GX 13+1 does not show the same narrow component of the low ionisation iron line in its {\it Chandra} spectrum \citep{Ueda2004} as is seen here in 2S0921.






\section{{\it XRISM}/Resolve simulations}

XRISM/Resolve is a microcalorimeter expected to launch later this year, which should give a resolution of $\sim 5$~eV at $\sim 6$~keV, enabling the dynamics of highly ionised gas to be resolved \citep{Tashiro2020}.
We simulate our best fit model to the {\it Chandra}/HETGS spectrum shown in Fig.~\ref{fig:heg_meg_rhd} and Fig.~\ref{fig:rhd_zoom} through the  {\it XRISM}/Resolve response.
We use the same 
exposure time ($7.5\times 10^{4} {\rm s}$) as used for {\it Chandra}, and rebin the resulting spectrum to 20 counts per bin (Fig.\ref{fig:xrism}).
We fit this using the same continuum/reflection models as in section 3, with two {\tt Pion} components to model the emission lines (parameters detailed in Table A3). Fig.\ref{fig:xrism} shows that there is a  weak but clear residual remaining picking out the absorption feature on the blue wing of the emission lines from Fe\,{\sc xxv} with  an equivalent width of around -5~eV.
This feature cannot be observed in either {\it XMM-Newton} or {\it Chandra} due to insufficient energy resolution.
This demonstrates that the wind signature in this ADC source is detectable with calorimeter resolution. 
{\it XRISM} will give a definitive test as to whether the ADC material originates in a wind.

This simulation also shows that {\it XRISM}/Resolve will show the detailed profile of the (semi-) neutral iron line(s). The line is well fitted with a 
single Gaussian with resolved width of 
$\sigma_{\rm line} \sim 0.01~{\rm keV}$, which is an order of magnitude lower than suggested from the Suzaku data analysis of \citet{Yoneyama2022}. 
Thus, {\it XRISM}/Resolve will also resolve the line width of the neutral iron line more accurately, though we also note that the current intrinsic binning of the  
{\tt xillverNS} is 20~eV, insufficient for {\rm XRISM}/Resolve data.

\section{Conclusions}

We analyse the spectrum ADC source 2S 0921-630 taken by the {\it XMM-Newton} and {\it Chandra}.
Both spectra are well described by a soft state neutron star continuum and its reflection from the outer disc plus emission lines from a complex (at least two component) photo-ionised plasma. The continuum shape is 
similar to the soft state of Atoll-sources, but with much lower luminosity. This all supports the identification of this source as an ADC, 
where the intrinsic luminosity of this source
$\gtrsim 0.1L_{\rm Edd}$, is seen only via scattering in diffuse gas above the accretion disc. 

We replace the two {\it ad hoc} photoionised emission components with the diffuse emission predicted from scattering and emission in a  thermal-radiative wind \citep{Tomaru2020b}.
This gives a good fit to all the {\it Chandra}/HETGS line emission in the 0.7-8~keV bandpass. This strongly supports the identification of the thermal-radiative wind as the origin of the scattering material in ADC sources. We show that a high-resolution calorimeter observation could directly show this through observing the predicted blueshifted absorption in Fe {\sc xxv}.

\section*{Acknowledgements}

CD thanks
Yue (Cory) Zhao and Poshak Gandhi for the GAIA distance estimate. 
CD and RT acknowledge
the Science and Technology Facilities Council (STFC) through grant
ST/T000244/1 for support. CD thanks Kavli IPMU, University of Tokyo for visitor funding.
AT is supported by the KU-DREAM program at Kagoshima University.
Numerical computations were in part carried out on Cray XC50 at Center for Computational Astrophysics (CfCA), National Astronomical Observatory of Japan (NAOJ).
Numerical analyses were in part carried out on analysis servers at CfCA, NAOJ.

\section*{Data Availability}
The {\it Chandra} and {\it XMM-Newton} data are publicly available. 
 Access to the radiation hydrodynamic code is available on request from R.T. (ryota.tomaru@durham.ac.uk).
 Access to the radiation transfer code is available on request from H.O.(odaka@ess.sci.osaka-u.ac.jp).



\bibliographystyle{mnras}
\bibliography{library} 




\appendix

\section{Full spectral fitting parameters}

\begin{table*}
	\centering
	\caption{The fitting results for {\it XMM-Newton}/EPIC-PN.  
 Each line shows different models. From left to right, continuum+reflection with 2 zone emissions from photoionized plasma, continuum+reflection with 1 zone emission, continuum+reflection, and continuum.
 In the model of reflection {\tt xillverNS}, we fix the iron abundance as solar ($A_{\rm Fe} = 1$), the density as $\log N=15 [{\rm cm^{-3}}]$, the red-shift $z=0$, and the inclination as $i=82^\circ$. 
 Also, we only consider reflection component fixing by ${\rm relf\_frag}=-1$. 
	}
	\label{tab:fit_xmm}
	\begin{tabular}{ |c|c|c|c|c|c|c} 
		\hline
		models & parameters & 2-zone & 1-zone  & continuum+reflection & continuum & 2-zone (with RGS)\\
		\hline
{\tt tbabs}       & $N_{H} [10^{22}{\rm cm^{-2}}]$   &$0.15\pm0.01$        & $0.15\pm0.01$  & $0.196^{+0.009}_{-0.005}$ &$0.15\pm0.01$ &$0.140^{+0.006}_{-0.01}$\\ [1ex]
{\tt Pion emit 1} & $\log \xi [{\rm erg~cm~s^{-1}}]$ & $4.7^{+0.1}_{-0.2}$ & $3.92^{0.06}_{-0.08} $       & &&$4.9^{+0.7}_{-0.3}$  \\[1ex]
                  & $N_{H} [10^{22}{\rm cm^{-2}}]$   & $>70$       & $60^{+10}_{-20}$      & && $200^{+500}_{-90}$  \\[1ex]
                  & $v_{\rm mic} [{\rm km~s{^-1}}]$      & $130^{+70}_{-80}$      & $120^{+40}_{-30}$ &&& $140^{+90}_{-50}$            \\[1ex]
{\tt Pion emit 2} & $\log \xi [{\rm erg~cm~s^{-1}}]$ & $3.3\pm0.1 $ &       &    && $3.3^{+0.2}_{-0.1}$      \\[1ex]
                  & $N_{H} [10^{22}{\rm cm^{-2}}]$   & $8\pm3$                       &         & &&$9^{+9}_{-3}$\\[1ex]
                  & $v_{\rm mic} [{\rm km~s{^-1}}]$      &  $230^{+130}_{-30}$                      &         &  &&$400\pm 100$\\[1ex]
{\tt nthcomp}     & $\Gamma$                         &$ 1.51^{+0.07}_{-0.04} $     & $1.53^{+0.06}_{-0.05}$  & $1.81^{+0.07}_{-0.05}$ &$1.45\pm 0.02$ &$1.51^{+0.02}_{-0.05}$\\[1ex]
                  & $kT_{\rm e} [{\rm keV}]$         &$ 2.1^{+0.2}_{-0.1}$ & $2.14^{+0.09}_{-0.08} $            & $2.52^{+0.08}_{-0.05}$ & $2.30^{+0.05}_{-0.06}$ & $2.06^{+0.2}_{-0.06} $\\[1ex]
                  & $kT_{\rm bb} [{\rm keV}]$        &$ 0.61^{+0.09}_{-0.08} $.      & $0.63\pm0.08 $             &$0.73^{+0.04}_{-0.03}$ & $0.48^{0.05}_{-0.08}$ & $0.62^{+0.05}_{-0.1}$\\[1ex]
                  & norm [$\times 10^{-3}$]          &$ 3.0\pm 0.5 $          & $2.9 ^{+0.4}_{-0.3}$               &$2.6^{+0.2}_{-0.1}$& $3.9^{+0.7}_{-0.4}$ &$2.9^{+0.5}_{-0.1}$\\[1ex] 
{\tt diskbb} & $kT [{\rm keV}]$ & $0.36\pm0.04$ & $0.38\pm 0.03$ & $0.27^{+0.01}_{-0.02}$& $0.34^{+0.03}_{-0.04}$ & $0.37^{+0.04}_{0.02}$ \\[1ex] 
             & norm  &  $40^{20}_{10}$ & $34 ^{+10}_{-9}$ & $43^{+10}_{-4}$& $40^{+30}_{-10}$ &$30\pm10$ \\[1ex] 
             
{\tt xillverNS} & $\log \xi$ &  <2 & $2.03^{+0.04}_{-0.08}$ & $3.00\pm 0.02$ & & <2 \\[1ex] 
                &  norm [$\times 10^{-5}$] &$14^{+4}_{-8}$ &$14^{+4}_{-3}$ &$18^{+2}_{-1}$ & & $ 15^{+3}_{-2}$   \\[1ex]
		\hline
Fit statistic     & $\chi^2/\nu $                    & 1967/1865  &2105/1868 &2304/1871 &2874/1873 &4159/3216\\
		\hline 
	\end{tabular}
\end{table*}

\begin{table*}
	\centering
	\caption{The fitting results for {\it Chandra}/HETGS.
 Each line shows different models. From left to right, continuum+reflection with 2 zone emissions from photoionized plasma, 1-zone of that, and  that from MCRT.
 We use the same parameters of continuums derived from {\it XMM-Newton}/EPIC-PN except for normalization. 
	}
	\label{tab:fit_chandra}
	\begin{tabular}{ |c|c|c|c|c|} 
		\hline
		models & parameters & 2 zone & 1 zone & RHDwind \\
		\hline
{\tt tbabs}       & $N_{H} [10^{22}{\rm cm^{-2}}]$   &$0.15$ (fixed)        & $0.15$ (fixed) & 0.15 (fixed)\\ [1ex]
{\tt Pion emit 1} & $\log \xi [{\rm erg~cm~s^{-1}}]$ & $4.7$ (fixed) & $3.7\pm 0.2 $        \\[1ex]
                  & $N_{H} [10^{22}{\rm cm^{-2}}]$   & $120$ (fixed)       & $40^{+40}_{-20}$  \\[1ex]
                  & $v_{\rm mic} [{\rm km~s{^-1}}]$      & $130$ (fixed)      & $210^{+40}_{-50}$            \\[1ex]
{\tt Pion emit 2} & $\log \xi [{\rm erg~cm~s^{-1}}]$ & $3.3$ (fixed)   \\[1ex]
                  & $N_{H} [10^{22}{\rm cm^{-2}}]$   & $8$   (fixed)  \\[1ex]
                  & $v_{\rm mic} [{\rm km~s{^-1}}]$      &  $230$ (fixed)                     &          \\[1ex]
{\tt RHDwind} & $\cos\theta$ & & & $0.301^{0.03}_{-0.007}$ \\[1ex]
{\tt nthcomp}     & $\Gamma$                         &$ 1.51$ (fixed)    & $1.53$ (fixed) &1.51 (fixed) \\[1ex]
                  & $kT_{\rm e} [{\rm keV}]$         &$ 2.1$ (fixed) & $2.1 $ (fixed) & 2.1 (fixed)  \\[1ex]
                  & $kT_{\rm bb} [{\rm keV}]$        &$ 0.61 $ (fixed)     & $0.63$ (fixed) & 0.61 (fixed)\\[1ex]
                  & norm [$\times 10^{-3}$]          &$ 1.59\pm 0.03 $          & $1.51 \pm 0.03 $ & $30.0 \pm 0.6$               \\[1ex] 
{\tt diskbb} & $kT [{\rm keV}]$ & $0.36$ (fixed)& $0.38$ (fixed) & 0.36 (fixed) \\[1ex] 
             & norm  &  $15\pm 1 $ & $14.1\pm 0.9$ & $270 \pm 20$ \\[1ex] 
{\tt xillverNS} & $\log \xi$ &  2 (fixed) & $2.0$ (fixed) & 2.0 (fixed)  \\[1ex] 
                &  norm [$\times 10^{-5}$] &$15 \pm 5 $ &$13\pm 5$ &$280 \pm 90$  \\[1ex]
		\hline
Fit statistic     & $\chi^2/\nu $                    & 1204/1482 &1206/1479 &1182/1481 \\
		\hline 
	\end{tabular}
\end{table*}

\begin{table*}
	\centering
	\caption{The fitting results for simulated {\it XRISM} observation 
 Here we use redshift as a free parameter.
	}
	\label{tab:fit_xrism}
	\begin{tabular}{ |c|c|c} 
		\hline
		models & parameters & 2 zone  \\
		\hline
{\tt tbabs}       & $N_{H} [10^{22}{\rm cm^{-2}}]$   &$0.25^{0.07}_{-0.04}$\\ [1ex]
{\tt Pion emit 1} & $\log \xi [{\rm erg~cm~s^{-1}}]$ & $5.4^{+0.3}_{-0.2}$  \\[1ex]
                  & $N_{H} [10^{22}{\rm cm^{-2}}]$   & $100^{+300}_{-50}$  \\[1ex]
                  & $v_{\rm mic} [{\rm km~s{^-1}}]$      & $180{+50}_{-60}$ \\[1ex]
                  &$z [ 10^{-4}]$ & $-1\pm2$ \\ [1ex]
{\tt Pion emit 2} & $\log \xi [{\rm erg~cm~s^{-1}}]$ & $3.72^{+0.05}_{-0.03}$ \\[1ex]
                  & $N_{H} [10^{22}{\rm cm^{-2}}]$   & $37^{+7}_{-6}$    \\[1ex]
                  & $v_{\rm mic} [{\rm km~s{^-1}}]$      &  $220 \pm 10$   \\[1ex]
                  & $z [10^{-4}$]& $6.1 \pm 0.6 $\\ [1ex]
{\tt nthcomp}     & $\Gamma$                         &$ 1.46^{+0.02}_{-0.03}$    \\[1ex]
                  & $kT_{\rm e} [{\rm keV}]$         &$ 1.93 ^{+0.04}_{-0.02}$  \\[1ex]
                  & $kT_{\rm bb} [{\rm keV}]$        &$ 0.43^{+0.07}_{-0.09} $   \\[1ex]
                  & norm [$\times 10^{-3}$]          &$ 2.2^{0.7}_{0.3} $           \\[1ex] 
{\tt diskbb} & $kT [{\rm keV}]$ & $0.24^{+0.03}_{-0.04}$  \\[1ex] 
             & norm  &  $150^{+230}_{-80} $  \\[1ex] 
{\tt xillverNS} & $\log \xi$ &  $1.8\pm 0.1$ \\[1ex] 
                &  norm [$\times 10^{-5}$] &$15 \pm 2 $  \\[1ex]
		\hline
Fit statistic     & $\chi^2/\nu $                    & 7021/6627  \\
		\hline 
	\end{tabular}
\end{table*}


\bsp	
\label{lastpage}
\end{document}